\documentclass[12pt]{amsart} 
\usepackage{euler}
\usepackage{euscript}
\usepackage{multicol} 
\usepackage{amssymb}
\usepackage{amsmath} 
\usepackage{times} 
\newcommand{\im}{\mbox{\upshape Im\ }} 
\newcommand{\re}{\mbox{\upshape Re\ }}
\newcommand{\tr}{\mbox{\upshape tr\ }}
\newtheorem{theorem}{Theorem}[section] 
\numberwithin{equation}{section}
\begin{document}
\title[Lieb-Thirring inequalities]{Sharp Lieb-Thirring Inequalities in
High Dimensions}
\author[A. Laptev and T. Weidl]{Ari
  Laptev$^1$ and Timo Weidl$^{1,2}$}
\begin{abstract}
We show how a matrix version of the Buslaev-Faddeev-Zakharov 
trace formulae
for a one-dimensional
Schr\"odinger operator leads to Lieb-Thirring inequalities
with sharp constants $L^{\mbox{\footnotesize\upshape cl}}_{\gamma,d}$  
with $\gamma\ge 3/2$ and arbitrary $d\ge 1$.
\end{abstract} 

\email{laptev@math.kth.se , weidl@math.kth.se} 
\maketitle

{
\setcounter{section}{-1}

\section{Introduction}

Let us consider a  Schr\"odinger operator in $L^2({\Bbb R}^d)$
\begin{equation}\label{schrOp}
 -\Delta + V\,,
\end{equation}
where $V$ is a real-valued function. 
In \cite{LT} Lieb and Thirring proved that if 
$\gamma>\max(0,1-d/2)$, then  there exist universal constants
$L_{\gamma,d}$
satisfying\footnote{Here and below we use the notion
$2x_-:=|x|-x$
for the negative part of variables, functions, Hermitian matrices 
or self-adjoint operators.
}
\begin{equation}
  \label{eq:LiebTh}
  \tr (-\Delta + V)_-^\gamma \leq L_{\gamma,d} \,
\int_{{\Bbb R}^d}
V_-^{\gamma+\frac{d}{2}}(x)\,dx\,.
\end{equation}
In the critical case $d\geq 3$ and $\gamma=0$ the
bound~\eqref{eq:LiebTh} 
is known 
as the Cwikel-Lieb-Rozenblum (CLR) inequality, see \cite{C,L1,R}
and also \cite{Con,LY}.  For the remaining case $d=1$, 
$\gamma=1/2$ the estimate \eqref{eq:LiebTh} has been verified
in \cite{W1}, see also \cite{HLT}.
On the other hand
it is known that  \eqref{eq:LiebTh}
fails for   $\gamma=0$ if $d=2$ and  
for $0\leq \gamma<1/2$ if $d=1$.
 
If $V\in L^{\gamma+\frac{d}{2}}({\Bbb R}^d)$, 
then the inequalities  \eqref{eq:LiebTh} are accompanied
by the Weyl type asymptotic formula
\begin{align}\notag
\lim_{\alpha\to+\infty}
\frac{1}
{\alpha^{\gamma+\frac{d}{2}}}\,\tr (-\Delta+\alpha V)^\gamma_- &= 
\lim_{\alpha\to+\infty}
\frac{1}{\alpha^{\gamma+\frac{d}{2}}}
\iint_{{\Bbb R}^d\times{\Bbb R}^d}
    (|\xi|^2+\alpha V)_-^\gamma\,\frac{dxd\xi}{(2\pi)^{d}}\\
\label{asyformu}
&=L^{\mbox{\footnotesize\upshape cl}}_{\gamma,d}
\int_{{\Bbb R}^d} V_-^{\gamma+\frac{d}{2}}\,dx\,,
\end{align}
where the so-called classical constant
$L^{\mbox{\footnotesize\upshape cl}}_{\gamma,d}$ 
is defined by
\begin{equation}\label{Lgd}
L^{\mbox{\footnotesize\upshape cl}}_{\gamma,d}= (2\pi)^{-d} \int_{{\Bbb
R}^d} (|\xi|-1)_-^\gamma
\,d\xi =
\frac{\Gamma(\gamma+1)}
{2^d\pi^{d/2}\Gamma(\gamma+\frac{d}{2}+1)}\,,
\quad \gamma\geq 0\,.
\end{equation}
It is interesting to compare the value of the sharp constant
$L_{\gamma,d}$ in~\eqref{eq:LiebTh} and the value of 
$L^{\mbox{\footnotesize\upshape cl}}_{\gamma,d}$. In particular,
the  asymptotic formula~\eqref{asyformu} implies that 
\begin{equation}\label{LLdg0}
L^{\mbox{\footnotesize\upshape cl}}_{\gamma,d}\leq L_{\gamma,d}
\end{equation}
for all $d$ and $\gamma$ whenever \eqref{eq:LiebTh} holds.  Moreover, in
\cite{AL} it has been shown, that for a fixed $d$ the ratio 
$L_{\gamma,d}/L^{\mbox{\footnotesize\upshape cl}}_{\gamma,d}$
is a monotone non-increasing function of $\gamma$.
In conjunction with the Buslaev-Faddeev-Zakharov trace formulae 
\cite{BF,FadZ} 
one obtains \cite{LT}
\begin{equation}\label{L=Lcl}
L_{\gamma,d} = L^{\mbox{\footnotesize\upshape cl}}_{\gamma,d}
\end{equation}
for
\begin{equation}\label{L=Lcl1}
d=1\qquad\mbox{and}\qquad \gamma\geq 3/2 \,.
\end{equation}
On the other hand one knows that
\begin{equation*}
L^{\mbox{\footnotesize\upshape cl}}_{\gamma,d}< L_{\gamma,d}
\end{equation*}
if $d=1$ and $1/2\leq \gamma<3/2$ (see \cite{LT}) or 
$\gamma<1$ and $d\in{\Bbb N}$ (see \cite{RoHe}). 

Up to now~\eqref{L=Lcl1}
was the only case 
where \eqref{L=Lcl} was known to be true for general classes 
of potentials $V\in L^{\gamma+\frac{d}{2}}$. 
Notice, however, that \eqref{L=Lcl} has been proven
for various {\em subclasses} of potentials.
If, for example, 
$\Omega\subset{\Bbb R}^d$ is a domain of finite measure, 
\begin{equation*}
V(x) = \begin{cases}-\alpha\quad&\mbox{as}\quad x\in\Omega\\
\infty\quad&\mbox{as}\quad x\in {\Bbb R}^d\setminus\Omega
\end{cases}\,, 
\end{equation*}
then 
the equality \eqref{L=Lcl} with $\gamma=0$  can be identified with the 
P\'olya conjecture on the number of the eigenvalues less than $\alpha$
for the Dirichlet Laplacian in $\Omega$. It holds true for 
tiling domains  \cite{P} and has been   
justified in \cite{Lap1} for certain domains 
of product structure  by using the method 
of ``lifting'' with respect to the dimension $d$ which is also 
one of the main ideas of this paper. 
If $\gamma\geq 1$, then 
\eqref{L=Lcl} is true for arbitrary $\Omega$.  This
is a simple corollary of the Berezin-Lieb inequality
(see \cite{Bz}, \S 5; \cite{L2} and also \cite{LS}).
This approach has been extended in \cite{Lap1} to 
the Dirichlet boundary value problems for 
matrices of pseudodifferential operators in ${\Bbb R^d}$ with 
constant coefficients.
The Berezin-Lieb inequality was also used in \cite{Lap2}
in order to improve the Lieb constant \cite{L1} in the CLR inequality 
for the subclass of Schr\"odinger operators whose potentials are 
equal to the characteristic functions of sets of finite measure. 

Another example is given in \cite{B}, where the identity \eqref{L=Lcl}
with $\gamma\geq 1$ and $d\in{\Bbb N}$
has been verified for a class of quadratic potentials.

We note, that
with the exception of~\eqref{L=Lcl1}, the sharp value of $L_{\gamma,d}$
has been recently found in \cite{HLT}, where it was proved that  for
$d=1$ and $\gamma=1/2$ 
\begin{equation*}
L_{1/2,1}=2L^{\mbox{\footnotesize\upshape cl}}_{1/2,1}=1/2\,.
\end{equation*}
In particular, in higher dimensions $d\geq 2$ 
the sharp values of 
the constants $L_{\gamma,d}$ have been unknown.

The main purpose of this paper is to verify \eqref{L=Lcl} 
for any $\gamma\geq 3/2$, $d\in{\Bbb N}$
and any $V\in L^{\gamma+\frac{d}{2}}({\Bbb R}^d)$.

In fact, this result is obtained for infinite-dimensional
systems of Schr\"odinger equations. 
Let ${\boldsymbol G}$ be a separable Hilbert space, let
${\boldsymbol 1}_{\boldsymbol G}$ be the identity operator on
${\boldsymbol G}$ and
consider 
\begin{equation}
-\Delta \otimes {\boldsymbol 1}_{\boldsymbol G}+V(x)\,,\qquad x\in{\Bbb
R}^d\,,
\end{equation}
in $L^2({\Bbb R}^d,{\boldsymbol G})$.
Here $V(x)$ is a family of 
self-adjoint non-positive
operators in ${\boldsymbol G}$, such that 
$\tr V\in L^{\gamma+\frac{d}{2}}({\Bbb R}^d)$.
Then we prove that
\begin{equation}\label{ssyss}
\tr\left(-\Delta\otimes {\boldsymbol 1}_{\boldsymbol
G}+V(x)\right)_-^\gamma
\leq L^{\mbox{\footnotesize\upshape cl}}_{\gamma,d}
\int_{{\Bbb R}^d}
\tr V_-^{\gamma+\frac{d}{2}}(x)\,dx
\end{equation}
for all $\gamma\geq 3/2$ and $d\geq 1$.
The inequality \eqref{ssyss} can  be extended 
to magnetic Schr\"odinger operators and we apply it to 
the Pauli operator.

We shall first deduce~\eqref{ssyss} for $d=1$, $\gamma=3/2$ and 
${\boldsymbol G}={\Bbb C}^n$
from the appropriate trace formula~\eqref{final2} for a finite system
of one-dimensional Schr\"odinger operators. 
In the scalar case 
these trace identities
are known as Buslaev-Faddeev-Zakharov formulae \cite{BF,FadZ}.
The matrix case
can be handled in a similar way as in the scalar case (see \cite{FadZ})
However, we give rather complete proofs of the 
corresponding statements in section 1, since we were unable to find 
the necessary formula~\eqref{final2} in the 
numerous papers devoted to this subject.

Note that we discuss trace formulae only as a technical tool
in order  to establish bounds on the negative spectrum. 
We therefore develop the theory 
of trace identities only as far as it is necessary 
for our own purpose.

In section 2 we 
extend the results of section 1 to the Schr\"odinger operator in 
$L^2({\Bbb R}^1,{\boldsymbol G})$. Applying  
a ``lifting'' argument with respect to 
dimension as used in \cite{GGM}
and \cite{Lap1}, we obtain in section 3
the main results of this paper.

Finally we would like to notice that the combination of the 
results of this paper and  
the equality $L_{1/2,1}=1/2$ discovered in
\cite{HLT} has lead  to new bounds on the Lieb-Thirring 
constants in \cite{HLW} which improve the corresponding bound 
obtained in \cite{Bla} and \cite{L3}.  

\section{Trace formulae for elliptic systems}

\subsection{Jost Functions}
Let ${\boldsymbol 0}$ and ${\boldsymbol 1}$ be the 
zero and the identity operator on ${\Bbb C}^n$. 
We consider the system of ordinary differential equations
\begin{equation}\label{y}
-\left(\frac{d^2}{dx^2}\otimes {\boldsymbol
1}\right)y(x)+V(x)y(x)=k^2y(x)\,,
\qquad x\in{\Bbb R}\,,
\end{equation}
where $V$ is a compactly supported, smooth 
(not necessary sign definite) Hermitian matrix-valued
function. Define  
\begin{equation*}
x_{\min}:=\min \mbox{supp}\,V
\quad\mbox{and}\quad 
x_{\max}:=\max\mbox{supp}\,V\,.
\end{equation*}
Then for any $k\in{\Bbb C}\backslash\{0\}$  there exist 
unique $n\times n$ matrix-solutions $F(x,k)$ and $G(x,k)$ of the
equations
\begin{align}
\label{Feq}
-F^{\prime\prime}_{xx}(x,k)+VF(x,k)&=k^2F(x,k)\,,\\
\label{Geq}
-G^{\prime\prime}_{xx}(x,k)+VG(x,k)&=k^2G(x,k)\,,
\end{align}
satisfying
\begin{alignat}{4}
\label{asF}
F(x,k)&=e^{ikx} {\boldsymbol 1} \quad&\mbox{as}\quad x&\geq
x_{\max}\,,\\
\label{asG}
G(x,k)&=e^{-ikx}{\boldsymbol 1} \quad&\mbox{as}\quad x&\leq x_{\min}\,.
\end{alignat}
If $k\in{\Bbb C}\setminus\{0\}$, then the pairs 
of matrices $F(x,k)$, $F(x,-k)$ and $G(x,k)$, $G(x,-k)$ form
full systems of independent solutions of \eqref{y}.
Hence the matrix
$F(x,k)$ can be expressed as a linear combination of $G(x,k)$ and
$G(x,-k)$
\begin{equation}\label{1}
F(x,k)=G(x,k)B(k)+G(x,-k)A(k)
\end{equation}
and vice versa:
\begin{equation}\label{2}
G(x,k)=F(x,k)\beta(k)+F(x,-k)\alpha(k)\,.
\end{equation}

\subsection{Basic properties of the matrices $A(k),$ $B(k),$ $\alpha(k)$
  and $\beta(k)$ for real $k$.}

Throughout this subsection we assume that $k\in{\Bbb R}\backslash\{0\}$.
Consider the Wronskian type matrix function
\begin{equation*}
W_1[F,G](x,k)=G^*(x,k)F^\prime_x(x,k)-(G^\prime_x(x,k))^*F(x,k)\,.
\end{equation*}
Then by \eqref{Feq} and \eqref{Geq} 
for $k\in{\Bbb R}$ we find that
\begin{equation*}
  \frac{d}{dx}W_1[F,G](x,k)= 
G^*(x,k)F^{\prime\prime}_x(x,k)-(G^{\prime\prime}_x(x,k))^*F(x,k)={\boldsymbol
0}\,.
\end{equation*}
Note that for $x\leq x_{\min}$ by~\eqref{1} we have
\begin{align*}
  W_1[F,G](x,k)&=\left[G^*(x,k)G^\prime_x(x,k)-
    (G^{\prime}_x(x,k))^*G(x,k)\right]B(k)\\ 
&\quad+\left[G^*(x,k)G^\prime_x(x,-k)-(G^{\prime}_x(x,k))^*G(x,-k)\right]A(k)\\
  &=-2ikB(k)\,,
\end{align*}
while for $x\geq x_{\max}$ by~\eqref{2} we find
\begin{align*}
  W_1[F,G](x,k)&= 
\beta^*(k)\left[F^*(x,k)F^\prime_x(x,k)-(F^\prime_x(x,k))^*F(x,k)\right]\\
  &\quad+ \alpha^*(k)
  \left[F^*(x,-k)F^\prime_x(x,k)-(F^\prime_x(x,-k))^*F(x,k)\right]\\
  &=2ik\beta^*(k)\,.
\end{align*}
This allows us to conclude that
\begin{equation}\label{B}
\beta^*(k)=-B(k)\,.
\end{equation}
Similarly, for the matrix-valued function
\begin{equation*}
W_2[F,G](x,k)=G^*(x,k)F^\prime_x(x,-k)-(G^\prime_x(x,k))^*F(x,-k)
\end{equation*}
we have  $\frac{d}{dx}W_2[F,G](x,k)={\boldsymbol 0}$ and 
\begin{alignat*}{4}
W_2[F,G](x,k)&=-2ikA(-k)\qquad&\mbox{as}\qquad x&\leq x_{\min}\,,\\
W_2[F,G](x,k)&=-2ik\alpha^*(k)\qquad&\mbox{as}\qquad x&\geq x_{\max}\,.
\end{alignat*}
Thus,
\begin{equation}\label{A}
A(-k)=\alpha^*(k)\,.
\end{equation}
Inserting~\eqref{1} into~\eqref{2} and making use 
of~\eqref{B},~\eqref{A} we obtain
\begin{align}\label{C}
  G(x,k)
  &=G(x,k)\left[B(k)\beta(k)+A(-k)\alpha(k)\right]\\
\notag
  &\quad +G(x,-k)\left[A(k)\beta(k)+B(-k)\alpha(k)\right]\,,
\end{align}
and thus
\begin{align}\label{D}
  A(-k)A^*(-k)-B(k)B^*(k)&={\boldsymbol 1}\,,\\
\label{D1}
  B(-k)A^*(-k)-A(k)B^*(k)&={\boldsymbol 0}\,.
\end{align}
In particular, this implies 
\begin{equation}\label{E}
|\det A(k)|^2=\det A(k)\det A^*(k)=
\det ({\boldsymbol 1}+B(-k)B^*(-k))\geq 1 
\end{equation}
for all $k\in{\Bbb R}\backslash\{0\}$.

\subsection{Associated Volterra equations and auxiliary estimates}

Next we derive estimates for the fundamental solutions of~\eqref{y} for
$\im k\geq 0$. Note first that
the matrices $F(x,k)$ and $G(x,k)$ are
solutions of the integral equations
\begin{align}\label{FF}
F(x,k)&=e^{ikx}{\boldsymbol 1}-\int_x^\infty 
k^{-1}\sin k(x-t) V(t)F(t,k)\,dt\,,\\
\label{GG}
G(x,k)&=e^{-ikx}{\boldsymbol 1}+\int_{-\infty}^x
k^{-1}\sin k(x-t) V(t)G(t,k)\,dt\,.
\end{align}
Put
\begin{equation*}
H(x,k)=e^{-ikx}F(x,k)-{\boldsymbol 1}\,.
\end{equation*}
Obviously, this matrix-valued function satisfies 
\begin{equation}
\label{HK0}
H(x,k)={\boldsymbol 0}\quad\mbox{for}\quad x\geq x_{\max}
\end{equation}
and 
\begin{equation}\label{H}
H(x,k)=\int_x^\infty K(x,t,k)\,dt + \int_x^\infty K(x,t,k)H(t,k)\,dt,
\end{equation}
where
\begin{equation}\label{KkK}
  K(x,t,k)=\frac{e^{2ik(t-x)}-1}{2ik}V(t)\,.  
\end{equation}
Note that
\begin{equation}\label{K}
\|K(x,t,k)\|\leq C_1(V,n)/(1+|k|)
\end{equation}
for all $k$ with $\im k\geq 0$ and all $k$ with $x_{\min}\leq x\leq t$.  
Here and below $\|\cdot\|$ denotes the norm of a matrix on ${\Bbb C}^n$.

Solving the Volterra equation \eqref{H} 
we obtain the convergent series 
\begin{equation*}
H(x,k)=\sum_{m=1}^\infty 
\underset{x\leq x_1\leq \dots\leq x_m}{\int\cdots\int}
\prod_{l=1}^m K(x_{l-1},x_l,k)\,dx_1\cdots dx_m\,.
\end{equation*}
%
%
%
{}From \eqref{K} we see that $|H(x,k)|\leq C_2(V)$
for all $x_{\min}\leq x\leq x_{\max}$.
Inserting this estimate back into~\eqref{H}, we conclude that the 
inequality
\begin{equation}
\label{HK1}\|H(x,k)\|\leq C_3(V,n) (1+|k|)^{-1}
\end{equation}
holds for all $x$ with $x_{\min}\leq x\leq x_{\max}$
and all $k$ with $\im k\geq 0$. 

\noindent
{\bfseries Remark 1.1.}
If we assume that $\im k \geq 0$ and $|k|\geq 1$,
then~\eqref{K} and therefore~\eqref{HK1}
holds true for all $x\in{\Bbb R}$.
\vspace{3mm}

It is not difficult to observe,
that $H(x,k)$ defined by~\eqref{H} is smooth in
\begin{equation*} 
(x,k)\in {\Bbb R}\times \{k\in {\Bbb C}: \im k\geq 0\}\,.
\end{equation*}
In particular, if we
differentiate~\eqref{H} with respect to $\overline{k}$
we find that
\begin{equation*}
\frac{\partial}{\partial\overline{k}}H(x,k)
= 
\int_x^\infty K(x,t,k)
\frac{\partial}{\partial\overline{k}}H(t,k)\,dt\,.
\end{equation*}
Since $\partial H(x,k)/\partial\overline{k}$ satisfies a
homogeneous
Volterra integral equation with the kernel \eqref{KkK},
we obtain $\partial H(x,k)/\partial\overline{k}\equiv 0$, 
and thus all the entries of the matrix 
$H(x,k)$ are analytic in $k$ for $\im k>0$.

\subsection{Further Estimates on $A(k)$ and $B(k)$.}
If we rewrite \eqref{FF} as follows
\begin{align}\label{Fk}
  F(x,k)&=e^{ikx}\left[{\boldsymbol 1}-\frac{1}{2ik}\int_x^\infty
V(t)\,dt - 
\frac{1}{2ik}
  \int_x^\infty V(t)H(t,k)\,dt\right]\\
\notag &\quad+\frac{e^{-ikx}}{2ik}\left[\int_x^\infty
e^{2ikt}V(t)\,dt+\int_x^\infty
  e^{2ikt}V(t)H(t,k)\,dt \right]\,,
\end{align}
then the expressions in the brackets in the r.h.s.\ do not depend on $x$
for
$x\leq x_{\min}$. Comparing \eqref{Fk} with~\eqref{1} we see that
\begin{align}\label{aa}
  A(k)&={\boldsymbol 1}-\frac{1}{2ik}\int_{-\infty}^{+\infty}V(t)\,dt -
\frac{1}{2ik}
  \int_{-\infty}^{+\infty}V(t)H(t,k)\,dt\,,\\
\label{bb}
B(k)&=\frac{1}{2ik}\int_{-\infty}^{+\infty}e^{2ikt}V(t)\,dt
+\frac{1}{2ik}\int_{-\infty}^{+\infty}e^{2ikt}V(t)H(t,k)\,dt\,.
\end{align}
For sufficiently large $|k|>C$  the smoothness of $V$ 
and~\eqref{HK1}  imply
\begin{align}
\label{Ak}
\left\|A(k)-{\boldsymbol
1}+\frac{1}{2ik}\int_{-\infty}^{+\infty}V(t)dt\right\|
&\leq C_4(V,n)|k|^{-2},\quad \im k\geq 0\,,\\
\label{Bk}
\left\|B(k)\right\| &\leq C_5(V,n)|k|^{-2}\,,\quad k\in {\Bbb R}\,.
\end{align}
In subsection 1.6
we shall see 
that  \eqref{Bk} can be improved so that
\begin{equation}\label{Bm}
B(k)=O(|k|^{-m})\quad\mbox{for all}\quad m\in{\Bbb N}\quad\mbox{as}\quad
k\to\pm\infty\,.
\end{equation}

\subsection{The matrix $A(k)$ for $\im k\geq 0$.}
First note that 
all entries of the matrix $A(k)$ are analytic in $k$ for $\im k> 0$
and continuous for $\im k\geq 0$, $k\neq 0$. 
This follows from
\eqref{aa} and the analyticity of $H(x,k)$. 
Fixing a sufficiently small $\epsilon>0$ and by using~\eqref{aa}
and~\eqref{HK1} we obtain
\begin{equation}\label{k00}
\|A(k)\|\leq C_6|k|^{-1}\qquad\mbox{as}\qquad |k|<\epsilon,\quad \im
k\geq
0.
\end{equation} 
Moreover, all the entries of $A(k)$ and thus the function $\det A(k)$
are analytic for $\im k>0$ and continuous for $\im k\geq 0$, $k\neq 0$.
Near the point $k=0$ we find
\begin{equation}\label{k01}
|\det A(k)|\leq C_7|k|^{-n}\qquad\mbox{as}\qquad |k|<\epsilon,\quad \im
k\geq
0.
\end{equation} 

Next let us describe the connection between the function
$\det A(k)$ and
the spectral properties of the 
self-adjoint problem \eqref{y} 
on $L^2({\Bbb R},{\Bbb C}^n)$.
Our assumptions on the matrix potential $V$ imply, 
that the operator on the l.h.s.\ of \eqref{y} 
has a discrete negative spectrum, which consists of 
finitely many negative eigenvalues 
$\lambda_l=(i\varkappa_l)^2$, $\varkappa_l>0$ of finite
multiplicities $m_l$. Obviously 
a solution $y(x)$ of \eqref{y} with $k=i\varkappa_l$ belongs to
$L^2({\Bbb R},{\Bbb C}^n)$, if and only if
\begin{alignat*}{4}
  y(x)&=G(x,i\varkappa_l)e^G_y \quad &\mbox{as}\quad x&\leq
x_{\min}\,,\\
  y(x)&=F(x,i\varkappa_l)e^F_y \quad &\mbox{as}\quad x&\geq x_{\max}\,,
\end{alignat*}
for some non-trivial vectors $e_G\,,e_F \in{\Bbb C}^n$. 
Linear independent solutions $y_1,\dots,y_{m_l}$
define linear independent vectors 
$e^G_{y_1},\dots,e^G_{y_{m_l}}$ and $e^F_{y_1},\dots,e^F_{y_{m_l}}$,
respectively. 
In view 
of~\eqref{1} we conclude that 
\begin{equation}
\dim\ker A(i\varkappa_l)=m_l\,.
\end{equation}
If we select an orthonormal basis in ${\Bbb C}^n$, such that
the first $m_l$ elements belong to $\ker A(i\varkappa_l)$,
we find that the first $m_l$ rows of $A(k)$ vanish as $k\to
i\varkappa_l$.
Since $\det A(k)$ does not depend on the choice of the orthonormal basis
and all entries of $A(k)$ are analytic, the function 
$\det A(k)$ has a zero of the order 
\begin{equation}\label{M<M}
m_l^\prime\geq m_l
\end{equation}
at $k=i\varkappa_l$, $\varkappa_l>0$. 
Moreover, if $\lambda=k^2$, $\im k>0$  is not an
eigenvalue of the problem \eqref{y}, then $\det A(k)\neq 0$.

In the remaining part of this subsection we prove that 
\begin{equation}\label{M=M}
m_l^\prime=m_l\,. 
\end{equation}
Let $g(x,y,k)$ be the Green function 
of the problem~\eqref{y}.
If $k^2<0$, $\im k>0$,
and $\det A(k)\neq 0$
it can be written as
\begin{equation*}
g(x,y,k)=
\begin{cases}
G(x,k)Z^-(y,k)\quad&\mbox{as}\quad y>x\\
-F(x,k)Z^+(y,k)\quad&\mbox{as}\quad y<x 
\end{cases} \,.
\end{equation*}
Here $Z^+(y,k)$ and $Z^-(y,k)$ are $n\times n$-matrices,
which are chosen such that 
\begin{align*}
\lim_{x=y-0} g(x,y;k)&=\lim_{x=y+0} g(x,y;k)\,,\\
\lim_{x=y-0} g^\prime_x(x,y;k)&=\lim_{x=y+0} g^\prime_x(x,y;k) +
{\boldsymbol 1}\,.
\end{align*}
These equations turn into
\begin{equation}\label{w-0}
W(y,k)
\begin{pmatrix}
Z^-(y,k)\\Z^+(y,k)
\end{pmatrix}
=
\begin{pmatrix}
{\boldsymbol 0}\\{\boldsymbol 1}
\end{pmatrix}
\,,\quad 
W(y,k)=
\begin{pmatrix}
G(y,k)&F(y,k)\\
G^\prime_y(y,k)&F^\prime_y(y,k)
\end{pmatrix}
\,.
\end{equation}
Since $\frac{\partial}{\partial y} \det W={\boldsymbol 0}$,
the determinant of $W$ is a constant with respect to $y$.
If  $y$ with $y<x_{\min}$, $\im k>0$, then
in view of~\eqref{1} and~\eqref{asG} we have
\begin{equation}\label{W}
W(y,k)=\begin{pmatrix}
e^{-iky}{\boldsymbol 1}&e^{-iky}B(k)+e^{iky}A(k)\\
-ike^{-iky}{\boldsymbol 1}&-ike^{-iky}B(k)+ike^{iky}A(k)
\end{pmatrix}
\,.
\end{equation}
Hence 
\begin{equation*}
\det W=(2ik)^n\det A(k)
\end{equation*}
and $W$ is invertible if and only if
$\det A(k)\neq 0$. 
{}From~\eqref{W} we see then, that 
for $y<x_{\min}$ the  entries $X_{ij}$ of 
\begin{equation}\label{w-1}
W^{-1}(y,k)=\begin{pmatrix}
X_{11}(y,k)&X_{12}(y,k)\\
X_{21}(y,k)&X_{22}(y,k)
\end{pmatrix}
\end{equation}
satisfy 
\begin{align*}
e^{-iky}X_{21}-ike^{-iky}X_{22}&={\boldsymbol 0}\,,\\
e^{-iky}(X_{21}-ikX_{22})B(k)+e^{iky}(X_{21}+ikX_{22})A(k)
&={\boldsymbol 1}\,.
\end{align*}
This gives
$X_{21}(y,k)=ikX_{22}(y,k)$,
and thus
\begin{equation*}
X_{22}(y,k)=(2ik)^{-1}e^{-iky}A^{-1}(k)\,.
\end{equation*}
In view of~\eqref{w-0} and~\eqref{w-1} we obtain
$Z^+(y,k)=X_{22}(y,k)$ and finally conclude that
\begin{equation}
g(x,y,k)=-(2ik)^{-1}A^{-1}(k)e^{ik(x-y)}
\quad\mbox{as}\quad y<x_{\min}<x_{\max}<x\,.
\end{equation}

If $k$ is in a sufficiently small neighbourhood of $i\varkappa_l$, 
the Green function 
$g(x,y,k)$ can be written as
\begin{equation*}
g(x,y,k)=
\frac{\sum_{r=1}^{m_l} \psi_r(x)\overline{\psi_r(y)}}
{(k-i\varkappa_l)(k+i\varkappa_l)}
+g_l(x,y,k)\,.
\end{equation*}
Here $g_l(x,y,k)$ is locally bounded and $\{\psi_r\}_{r=1}^{m_l}$ 
forms an orthonormal
eigenbasis corresponding to the eigenvalue $\lambda_l=-\varkappa_l^2$.
Hence,
\begin{align*}
\det X_{22}(y,k)&=(2ik)^{-n}e^{-inky}\det A^{-1}(k)\\
&=(-1)^n e^{-inkx}\det g(x,y,k)= O\left(|k-i\varkappa_l|^{-m_l}\right)
\end{align*}
as $k\to i\varkappa_l$. This implies that
$\det A(k)$ has a zero of the order 
\begin{equation*}\label{M>M}
m_l^\prime\leq m_l
\end{equation*}
at $k=i\varkappa_l$. 
Finally, the last inequality and~\eqref{M>M} imply~\eqref{M=M}.

\subsection{The matrix function $T(x,k)$.}

Consider the matrix function
\begin{equation}\label{T}
T(x,k)={\boldsymbol 1}+H(x,k)={\boldsymbol 1}+\int_x^\infty
K(x,t,k)T(t,k)\,dt\,.
\end{equation}
According to subsection 1.3 the matrix-valued function
$T(x,k)$ is smooth and uniformly bounded for
\begin{equation*}
(x,k)\in {\Bbb R}\times\left\{k\in{\Bbb C}: 
\im k\geq 0\quad\mbox{and}\quad |k|\geq 1\right\}\,.
\end{equation*}
Obviously $T(x,k)={\boldsymbol 1}$  for $x\geq x_{\max}$.
Integrating by parts in~\eqref{T} and using~\eqref{KkK} we obtain
\begin{equation}\label{dT}
\frac{d^l}{dx^l} T(x,k)=-\int_x^\infty e^{2ik(t-x)}
\frac{d^{l-1}}{dt^{l-1}}\left(V(t)T(t,k)\right)dt
\end{equation}
for all $l\in{\Bbb N}$. Since 
$\mbox{\upshape supp}\, V\subseteq [x_{\min},x_{\max}]$ we find
\begin{alignat}{6}
\label{prelT0}
d^lT(x,k)/dx^l
&=&\, &{\boldsymbol 0}&\quad&\mbox{as}&\quad 
x_{\max}&\leq x&\,&,&\\
\label{prelT1}
\left\|d^lT(x,k)/dx^l\right\|
&\leq&\, &C_8&\quad&\mbox{as}&\quad 
x_{\min}\leq x&\leq x_{\max}&\,&,&\\
\label{prelT2}
\left\|d^lT(x,k)/dx^l\right\|
&\leq&\, &C_9e^{2(x-x_{\min}){\footnotesize\im} k}&\quad&\mbox{as}&\quad 
x&\leq x_{\min}&\,&,&
\end{alignat}
for all $k$ with $\im k\geq 0$ and $|k|\geq 1$. The constants
$C_8$ and $C_9$ depend only upon $V$, $n$ and $l$.
If we integrate the r.h.s.\ of \eqref{dT} by parts,
then~\eqref{prelT1} and~\eqref{prelT2} imply
\begin{alignat}{6}
\label{finT1}
\left\|d^lT(x,k)/dx^l\right\|
&\leq&\, &\frac{C_{10}}{1+|k|}&\quad&\mbox{as}&\quad 
x_{\min}\leq x&\leq x_{\max}&\,&,&\\
\label{finT2}
\left\|d^lT(x,k)/dx^l\right\|
&\leq&\, &\frac{C_{11}}{1+|k|}e^{2(x-x_{\min}){\footnotesize\im}
k}&\quad&\mbox{as}&\quad 
x&\leq x_{\min}&\,&,&
\end{alignat}
for all $k$ with $\im k\geq 0$ and $|k|\geq 1$. The constants
$C_{10}$ and $C_{11}$ depend only upon $V$, $n$ and $l$.

In a similar way integrating
by parts in~\eqref{dT}, we 
obtain the asymptotical decompositions
\begin{align}\notag
  \frac{d^l}{dx^l}T(x,k)&=
  -\int_x^\infty e^{2ik(t-x)}
  \frac{d^{l-1}}{dt^{l-1}}(V(t)T(t,k))\,dt\\
  \notag &= \left\{\sum_{r=1}^q
    \frac{(-1)^{r+1}}{(2ik)^r}\frac{d^{r+l-2}}{dx^{r+l-2}}\right\}
  \left(V(x)T(x,k)\right)
  \\
  \notag &\qquad +(-1)^{q+1}\int_x^\infty \frac{e^{2ik(t-x)}}{(2ik)^q}
  \frac{d^{q+l-1}}{dt^{q+l-1}}\left(V(t)T(t,k)\right)\,dt\\
\label{dTas}
&=\left\{\sum_{r=1}^{q-1}
  \frac{(-1)^{r+1}}{(2ik)^r}\frac{d^{r+l-2}}{dx^{r+l-2}}\right\}
\left(V(x)T(x,k)\right)+R_{q,l}(x,k)
\end{align}
as $|k|\geq 1$, $\im k>0$. 
Here 
\begin{alignat}{6}
\label{Rm0}
R_{q,l}(x,k)
&=&\, &{\boldsymbol 0}&\quad&\mbox{as}&\quad 
x_{\max}&\leq x&\,&,&\\
\label{Rm1}
\left\|R_{q,l}(x,k)\right\|
&\leq&\, &C_{12}(1+|k|)^{-q}&\quad&\mbox{as}&\quad 
x_{\min}\leq x&\leq x_{\max}&\,&,&\\
\label{Rm2}
\left\|R_{q,l}(x,k)\right\|
&\leq&\, &\frac{C_{13}}{(1+|k|)^q}
e^{2(x-x_{\min}){\footnotesize\im} k}&\quad&\mbox{as}&\quad 
x&\leq x_{\min}&\,&.&
\end{alignat}
The constants $C_{12}$ and $C_{13}$ depend upon $V$, $n$, $l$ and $q$.

Since $d^lH/dx^l=d^lT/dx^l$ for all $l\in{\Bbb N}$,  
integration by parts in~\eqref{bb} and the 
inequalities~\eqref{prelT0},~\eqref{finT1}
and~\eqref{finT2} 
give~\eqref{Bm}.

\subsection{The matrix function $\sigma(x,k)$}
By using~\eqref{HK0},~\eqref{HK1} and Remark 1.1
for sufficiently large $|k|$, $\im k\geq 0$,
the matrix  $T(x,k)={\boldsymbol 1}+H(x,k)$ is invertible
for all $x\in{\Bbb R}$ and
\begin{equation}\label{T1}
\|T^{-1}(x,k)\|\leq C_{14}\quad\mbox{for all}\quad x\in{\Bbb R}\,,
\quad |k|>C_{15}\,,\quad \im k\geq 0\,,
\end{equation}
with sufficiently large constants $C_{14}=C_{14}(V,n)$ and
$C_{15}=C_{15}(V,n)$.
Hence, for sufficiently large $|k|$ with $\im k\geq 0$ the matrix
function
\begin{equation}
  \label{eq:sigma}
  \sigma(x,k)=\left[\frac{d}{dx}T(x,k)\right]T^{-1}(x,k)
\end{equation}
is well defined for all $x\in {\Bbb R}$. Liouville's formula
\begin{equation*}
\frac{d}{dx}(\ln\det
T(x,k))=\tr\left\{\left[\frac{d}{dx}T(x,k)\right]T^{-1}(x,k)\right\}
\end{equation*}
implies 
\begin{equation*}
\frac{d}{dx}\left(\ln\det e^{-ikx}F(x,k)\right)=\tr\sigma(x,k)\,.
\end{equation*}
Since $e^{-ikx}F(x,k)={\boldsymbol 1}$ as $x\geq x_{\max}$ and
\begin{equation*}
e^{-ikx}F(x,k)=e^{-2ikx}B(k)+A(k)=A(k)+o(1)
\end{equation*}
as $x\to-\infty$, $\im k\geq\epsilon> 0$, we finally conclude that
\begin{equation}\label{equ}
\ln\det A(k)=-\int_{-\infty}^{+\infty}\tr\sigma(x,k)\,dx\,,
\end{equation}
\begin{equation*}
|k|\geq C_{15}\,,\qquad \im k\geq \epsilon>0\,.
\end{equation*}
\vspace{3mm}

\noindent
{\bfseries Remark 1.2.} Formula~\eqref{equ} is a matrix version
of the corresponding
well-known identity for scalar Schr\"odinger operators 
(see e.g. \S 3 in \cite{FadZ}). 
\vspace{3mm}

\subsection{The asymptotical decomposition of $\sigma(x,k)$}
Next we shall
develop $\sigma(x,k)$ into an asymptotical series with respect 
to the inverse powers of $k$. 
For the sake of future references
we compute the first three terms, although we only need 
the second one in this paper.

If we apply~\eqref{dTas} with $q=2$, $l=1$ we find that
\begin{equation}\label{sigma1}
\sigma=\frac{1}{2ik}V+Q_2\,,\qquad Q_2=R_{2,1}T^{-1}\,,
\end{equation}
while~\eqref{dTas} with $q=4$, $l=1$ gives
\begin{align}\label{sigma2}
\sigma&=\frac{1}{(2ik)^3}
\left\{\frac{d^2V}{dx^2}+2\frac{dV}{dx}\sigma
+V\frac{d^2T}{dx^2}T^{-1}\right\}
\\
\notag
&\qquad-\frac{1}{(2ik)^2}
  \left\{\frac{dV}{dx}+V\sigma\right\}+ \frac{1}{2ik}V+R_{4,1}T^{-1}\,.
\end{align}
Inserting \eqref{sigma1} into \eqref{sigma2} we obtain
\begin{align}\label{sigma3}
  \sigma
=\frac{1}{2ik}V-\frac{1}{(2ik)^2}\frac{dV}{dx}
-\frac{1}{(2ik)^3}\left\{V^2-\frac{d^2V}{dx^2}\right\}+Q_4\,.
\end{align}
Finally, if we insert in a similar way~\eqref{sigma3} and~\eqref{dTas}
with $l=2$, $q=3$ as well as $l=3$, $q=2$ into~\eqref{dTas} 
with $l=1$ and $q=6$, we arrive at
\begin{align}\label{sigma6}
\sigma
&=(2ik)^{-1}V-
(2ik)^{-2}\frac{dV}{dx}+(2ik)^{-3}\left\{\frac{d^2V}{dx^2}-V^2\right\}\\
\notag
&\qquad-(2ik)^{-4}\left\{\frac{d^3V}{dx^3}
-2\frac{dV^2}{dx}\right\}\\
\notag
&\qquad+(2ik)^{-5}\left\{\frac{d^4V}{dx^4}-3\frac{d^2V^2}{dx^2}
+\left(\frac{dV}{dx}\right)^2 +2V^3\right\}
+Q_6\,.
\end{align}
As well as $R_{q,l}$ the terms $Q_2$, $Q_4$ and $Q_6$ satisfy 
the  inequalities of the type~\eqref{Rm0}  -- \eqref{Rm2}
with
$q=2$, $q=4$, and $q=6$, respectively. Then we conclude that
\begin{equation*}
  \int_{-\infty}^{+\infty} \tr Q_q(x,k)dx = O(|k|^{-q})\,,
\qquad q=2,4,6\,,
\end{equation*}
as $|k|\to\infty$ with $\im k\geq \epsilon>0$ and
thus, 
\begin{align}\label{trace}
  \int_{-\infty}^{+\infty} \tr\sigma(x,k)dx&=
  \frac{1}{2ik}\int_{-\infty}^{+\infty}\tr V\,dx
  -\frac{1}{(2ik)^3} \int_{-\infty}^{+\infty}\tr V^2\,dx\\
\notag
  &+\frac{1}{(2ik)^5} \int_{-\infty}^{+\infty}
\left[2\tr V^3+\tr\left(\frac{dV}{dx}\right)^2\right]\,dx
+O(|k|^{-6})
\end{align}
as $|k|\to\infty$ with $\im k\geq \epsilon>0$.

\subsection{The dispersion formula}

Let
\begin{equation*}
\{\lambda_l\}_{l=1}^N=\{(i\varkappa_l)^2\}_{l=1}^N, 
\qquad \varkappa_l>0,
\end{equation*}
be the finite set of the negative eigenvalues of \eqref{y}.  
Each eigenvalue occurs
in this set only once.  Let $m_l$ be the order of zero of
$\det
A(k)$ at the point $k=i\varkappa_l$,
which by section 1.5 equals the multiplicity of the corresponding
eigenvalue.
Then the arguments in section 1.5 imply that the function
\begin{equation}\label{M0}
M(k)= \ln \left\{\det A(k) \prod_{l=1}^N 
\left(\frac{k+i\varkappa_l}{k-i\varkappa_l}\right)^{m_l} \right\}
\end{equation}
is analytic for $\im k> 0$ and continuous up to the boundary
except
$k=0$, where it has at most a logarithmic singularity.
Moreover, the inequality \eqref{Ak} gives
\begin{equation*}
|M(k)|\leq C_2(V) |k|^{-1}
\end{equation*}
for all sufficiently large $|k|>C$, $\im k\geq 0$.
Hence, by
applying Cauchy's formula for large
semi-circles in the upper half-plane we obtain
\begin{equation*}
  \int_{-\infty}^{+\infty}\frac{M(z)\,dz}{z-k}=(2\pi i)M(k)\,,
  \qquad
  \int_{-\infty}^{+\infty}\frac{M(z)\,dz}{z-\overline{k}}=0
\end{equation*}
for arbitrary $k$ with $\im k>0$.
This implies
\begin{equation}\label{M}
M(k)=\frac{1}{\pi i} \int_{-\infty}^{+\infty}
\frac{\re M(z)}{z-k}\,dz\,,
\end{equation}
which by~\eqref{M0} is  equivalent to
\begin{equation}\label{AM}
\ln \det A(k) = 
\frac{1}{\pi i}
\int_{-\infty}^{+\infty}\frac{\ln |\det A(z)|\,dz}{z-k}
+ \sum_{l=1}^N 
m_l\ln \frac{k-i\varkappa_l}{k+i\varkappa_l} 
\end{equation}
for all $k$ with $\im k>0$.

\subsection{Trace formulae for elliptic systems}
Note that
\begin{align}
\notag
 \sum_{l=1}^Nm_l \ln \frac{k-i\varkappa_l}{k+i\varkappa_l} = 
\frac{2}{ik}\sum_{l=1}^Nm_l\varkappa_l
&-\frac{2}{3ik^3}\sum_{l=1}^Nm_l\varkappa^3_l\\
\label{ssums}
&+\frac{2}{5ik^5}\sum_{l=1}^Nm_l\varkappa^5_l
+O(|k|^{-6})
\end{align}
as $|k|\to\infty$, $\im k\geq \epsilon>0$.  On the other hand 
from~\eqref{E} and~\eqref{Bm} we have
\begin{equation*}
  \ln |\det A(z)|= 2^{-1}\ln |\det ({\boldsymbol 1}+B(-z)B^*(-z))| 
= O(|z|^{-m}), 
\qquad z\in{\Bbb R}\,,
\end{equation*}
as $|z|\to\infty$, for all $m\in{\Bbb N}$. 
Hence, the integral in~\eqref{AM} permits the
asymptotical decomposition
\begin{align}\notag
\int_{-\infty}^{+\infty}
\frac{\ln |\det A(z)|\,dz}{z-k}&=-\sum_{j=0}^m 
\frac{I_j}{k^{j+1}}+O(|k|^{m+1})\,,\\
\label{Aints}
I_j&=\int_{-\infty}^{+\infty}z^j \ln |\det A(z)|\,dz
\end{align}
as $|k|\to\infty$, $\im k\geq\epsilon>0$.

Combining~\eqref{ssums},~\eqref{Aints} with $m=5$ 
and~\eqref{trace} we obtain 
\begin{align}\label{final1}
  \frac{1}{4}\int\tr V\,dx &=
  \frac{I_0}{2\pi} 
  -\sum_{l=1}^Nm_l\varkappa_l\,,\\
\label{final2}
\frac{3}{16} \int\tr V^2\,dx &=
\frac{3I_2}{2\pi} +
\sum_{l=1}^Nm_l\varkappa^3_l\,,\\
\label{final3}
\frac{5}{32} \int\tr V^3\,dx
+\frac{5}{64} \int\tr \left(\frac{dV}{dx}\right)^2\,dx 
&=
\frac{5I_4}{2\pi} -
\sum_{l=1}^Nm_l\varkappa^5_l\,.
\end{align}
Finally we remark,
that in view of~\eqref{E} 
\begin{equation}\label{lj}
I_{j}\geq 0
\end{equation}
for all
even, non-negative integers $j$.

\section{Sharp Lieb-Thirring inequalities for second order 
one-dimensional Schr\"odinger
type  systems}

\subsection{A Lieb-Thirring estimate for finite systems}
Let us first consider the operator on the l.h.s.\ of~\eqref{y}
in $L^2({\Bbb R},{\Bbb C}^n)$ for some 
smooth, compactly supported Hermitian matrix potential $V$.
Preserving the notation of the previous section 
the bounds~\eqref{final2} and~\eqref{lj} imply
\begin{equation}
  \label{eq:LTH0}
  \tr\left(-\frac{d^2}{dx^2}\otimes{\boldsymbol 1}+V(x)\right)_-^{3/2} 
=\sum_l m_l\varkappa_l^3
\leq 
\frac{3}{16}\int \tr V^2 (x)\,dx\,.
\end{equation} 
By continuity~\eqref{eq:LTH0} extends to all
Hermitian matrix potentials, for which $\tr V^2$
is integrable.
Finally, a standard variational argument allows one
to replace $V$ by its negative part $V_-$:
\begin{equation}
  \label{eq:LTH}
  \tr\left(-\frac{d^2}{dx^2}\otimes{\boldsymbol 1}+V(x)\right)_-^{3/2}
\leq 
\frac{3}{16}\int \tr V^2_-(x)\,dx\,. 
\end{equation}
The constant in the r.h.s.\ of this inequality is sharp and coincides
with the
classical constant $L^{\mbox{\footnotesize\upshape cl}}_{3/2,1}$. 
In particular, this constant does not
depend on the internal dimension $n$ of the system.
 
\subsection{Operator-valued differential equations}

Let ${\boldsymbol G}$ be a separable Hilbert space
with the scalar product $<\cdot,\cdot>_{\boldsymbol G}$
and the norm $\|\cdot\|_{\boldsymbol G}$.
Let $H^1({\Bbb R},{\boldsymbol G})$ and $H^2({\Bbb R},{\boldsymbol G})$
be the Sobolev spaces of all functions
\begin{equation*} 
u(\cdot):{\Bbb R}\to{\boldsymbol G}\,, 
\end{equation*}
for which the respective norms
\begin{align*}
\|u\|^2_{H^1}&=
\int_{-\infty}^{+\infty} \left( \|u^\prime\|^2_{\boldsymbol G}
+
\|u\|^2_{\boldsymbol G}\right)\,dx\\
\|u\|^2_{H^2}&=
\int_{-\infty}^{+\infty} 
\left(
\|u^{\prime\prime}\|^2_{\boldsymbol G}
+
\|u\|^2_{\boldsymbol G}
\right)\,dx
\end{align*}
are finite. Finally, let ${\boldsymbol 1}_{\boldsymbol G}$
be the
identity operator on ${\boldsymbol G}$. Then the operator 
$-\frac{d^2}{dx^2}\otimes {\boldsymbol 1}_{\boldsymbol G}$ defined on
$H^2({\Bbb R},{\boldsymbol G})$ is self-adjoint in
$L^2({\Bbb R},{\boldsymbol G})$. It corresponds to the closed
quadratic form 
\begin{equation*}
h[u,u]=\int\left\|u^\prime\right\|^2_{\boldsymbol G}\,dx
\end{equation*}
with the form domain $H^1({\Bbb R},{\boldsymbol G})$.

Let ${\mathscr B}$ and ${\mathscr K}$ respectively be the spaces
of all bounded and compact linear operators on ${\boldsymbol G}$. 
Let $\|\cdot\|_{\mathscr B}$ 
denote the corresponding operator norm. Consider an operator-valued
function 
\begin{equation*}
W(\cdot):{\Bbb R}\to {\mathscr B}\,, 
\end{equation*}
for which $W(x)=(W(x))^*$, $x\in{\Bbb R}$ and 
$\|W(\cdot)\|_{\mathscr B} \in L^p({\Bbb R})$, $1<p<\infty$. 
Denote 
\begin{equation*}
w[u,u]=\int_{-\infty}^{+\infty}
\left<W(x)u(x),u(x)\right>_{\boldsymbol G}\,dx\,.
\end{equation*}
This form is well-defined on $H^1({\Bbb R},{\boldsymbol G})$ and
\begin{equation}\label{eq:16}
\left|w[u,u]\right| \leq C_{16} 
\left(
\int_{-\infty}^{+\infty} \|W(x)\|^p_{\mathscr B} dx
\right)^{1/p}\|u\|^2_{H^1}\,.
\end{equation}

The constant $C_{16}$ does not depend upon $W$ or $u$.
Moreover, for all $\epsilon>0$ there exists a finite constant
$C_{17}(\epsilon,W)$, such that 
\begin{equation}\label{eq:17}
\left|w[u,u]\right|
\leq \epsilon h[u,u] +
C_{17}(\epsilon,W)\int \|u\|^2_{\boldsymbol G}\,dx\,.
\end{equation}
Both~\eqref{eq:16} and~\eqref{eq:17} 
follow immediately from the corresponding inequalities
which hold in the scalar case.
Hence, the quadratic form
\begin{equation*}
h[u,u]+w[u,u]
\end{equation*} 
is semi-bounded from below and closed on $H^1({\Bbb R},{\boldsymbol
G})$.
It induces a self-adjoint semi-bounded operator
\begin{equation}\label{op}
Q=-\frac{d^2}{dx^2}\otimes {\boldsymbol 1}_{\boldsymbol G}+ W(x)
\end{equation}
on $L^2({\Bbb R},{\boldsymbol G})$. 

If in addition $W(x)\in{\mathscr K}$ for a.e. $x\in{\Bbb R}$,
then the form $w[\cdot,\cdot]$ is relative compact with
respect to the metric on $H^1({\Bbb R},{\boldsymbol G})$.
In order to prove this fact we introduce
the orthogonal projections $P_M$ on the linear span
of the first $M$ elements of some fixed orthonormal basis in
${\boldsymbol G}$.
As a consequence, the Birman-Schwinger principle implies,
that the negative spectrum of the operator $Q$ is discrete
and might accumulate only to zero. In other words, the operator
$Q_-$ is compact on $L^2({\Bbb R},{\boldsymbol G})$.

\subsection{A Lieb-Thirring estimate 
for operator-valued differential equations}

We shall prove the following Theorem: 

\begin{theorem}\label{t1} 
Let $W(x)$ be self-adjoint Hilbert-Schmidt operators on 
${\boldsymbol G}$ 
for a.e. $x\in{\Bbb R}$ and let $\tr W^2(\cdot)\in 
L^1({\Bbb R},{\boldsymbol G})$.
Then we have
\begin{equation}\label{infin}
\tr \left( -\frac{d^2}{dx^2}\otimes {\boldsymbol 1}_{\boldsymbol G}
+ W(x)\right)^{3/2}_-
\leq L^{\mbox{\footnotesize\upshape cl}}_{3/2,1} 
\int_{-\infty}^{+\infty} \tr W^2_-\,dx\,,
\end{equation}
where according to~\eqref{Lgd}  
it holds 
$L^{\mbox{\footnotesize\upshape cl}}_{3/2,1}=3/16$.
\end{theorem}

\noindent
{\bf Proof.} Assume that \eqref{infin} fails. Then there exists a
non-positive operator family $W$ satisfying 
$\tr W^2(\cdot)\in L^1({\Bbb R})$
and some sufficiently small $\epsilon>0$, such that
\begin{equation}\label{lhs}
\tr \chi_{\epsilon}^{3/2}(Q)  > \frac{3}{16} 
\int_{-\infty}^{+\infty} \tr W^2 \,dx\,.
\end{equation}
Here 
\begin{equation*}
\chi_{\epsilon}(Q)=-E_{(-\infty,-\epsilon)}(Q)Q\,,
\end{equation*}
with $E_{(-\infty,-\epsilon)}(Q)$ being 
the spectral projection of $Q$ onto 
the interval $(-\infty,-\epsilon)$.
Since $Q_-$ is compact, the operator 
$E_{(-\infty,-\epsilon)}(Q)$ is of a finite rank $n(\epsilon)$.

Fix some orthonormal basis in ${\boldsymbol G}$ and let ${\Bbb P}_M$ be
the
projection on the linear span of its first $M$ elements.  Consider the
auxiliary operators
\begin{equation*}
Q(M,\epsilon)=E_{(-\infty,-\epsilon)}(Q) (1(x)\otimes P_M)Q(1(x)\otimes
P_M)E_{(-\infty,-\epsilon)}(Q)\,.
\end{equation*}
Obviously we have $\mbox{rank}\,Q(M,\epsilon)\leq n(\epsilon)$
for all $M$. Since
$1(x)\otimes {\Bbb P}_M$ turns to
the identity operator on $L^2({\Bbb R},{\boldsymbol G})$ 
in the strong operator topology
as $M\to\infty$, then the operators
$Q(M,\epsilon)$ 
converge to $\chi_\epsilon(Q)$
in the $L^2({\Bbb R}{\boldsymbol G})$ operator norm, 
as $M\to\infty$ and
\begin{equation*}
\tr (Q(M,\epsilon))_-^{3/2}\to \tr \chi_\epsilon(Q)
\quad\mbox{as}\quad M\to \infty\,.
\end{equation*}
Thus,
\begin{equation}  
\label{eq:n3}
  \tr (Q(M,\epsilon))_-^{3/2} > \frac{3}{16} 
\int_{-\infty}^{+\infty} \tr W^2\,dx 
\end{equation}
for some sufficiently large $M$.
On the other hand, a standard variational argument implies
\begin{equation*}
  \tr (Q(M,\epsilon))_-^{3/2}\leq 
  \tr \left((1(x)\otimes P_M)Q(1(x)\otimes P_M)\right)_-^{3/2}\,.
\end{equation*}
Observe that the expression on the r.h.s.\ is nothing else but the Riesz
mean
of the order $\gamma=3/2$ of the negative eigenvalues of the $M\times
M$-system~\eqref{y}  with 
$V(x)=P_MW(x)P_M$.  
Thus, from~\eqref{eq:LTH} we obtain
\begin{equation*}
  \tr (Q(M,\epsilon))_-^{3/2}\leq \frac{3}{16}
\int \tr V^2(x)\,dx \leq \frac{3}{16}
\int \tr W^2(x)\,dx\,,
\end{equation*}
which contradicts~\eqref{eq:n3}.  
This completes the proof.

\subsection{Lieb-Thirring estimates for Riesz means of negative
eigenvalues of the order $\gamma\geq 3/2\,$}

We shall now suppose, that the non-positive
operator family $W(x)$ satisfies
\begin{equation}\label{WW}
\tr W^{\gamma+\frac{1}{2}}_-(x)\in L^1({\Bbb R})
\quad\mbox{for some}\quad \gamma>3/2\,.
\end{equation}
Let $dE_{(-\infty,\lambda)}(Q)$ be the spectral measure of the operator
$Q$.
Repeating the arguments of Aizenman and Lieb \cite{AL}, we find
\begin{align*}
  B\left(\gamma-\frac{3}{2},\frac{5}{2}\right) \tr Q_-^\gamma&=
  \tr
  \left\{\int_{-\infty}^0dE_{(-\infty,\lambda)}(Q)
\int_0^\infty t^{\gamma-\frac{5}{2}}(t+\lambda)_-^{3/2}\,dt\right\}\\
  &=\int_0^\infty  t^{\gamma-\frac{5}{2}}
  \tr (Q+t)_-^{3/2}\,dt\\
  &\leq \frac{3}{16} \int_0^\infty dt\, t^{\gamma-\frac{5}{2}}
  \int_{-\infty}^{+\infty}\tr(W(x)+t)_-^{2}\,dx\,,
\end{align*}
where 
%
$B(x,y)=\frac{\Gamma(x+y)}{\Gamma(x)\Gamma(y)}$
%
is the Beta function.
Let $-\mu_j(x)<0$ be the
negative eigenvalues of $W(x)$. Then
\begin{align*}
  \int_0^\infty dt\,t^{\gamma-\frac{5}{2}}
  \int_{-\infty}^{+\infty} &\tr(W(x)+t)_-^{2}\,dx\\
  &= \sum_{j=1}^\infty \int_{-\infty}^{+\infty} dx \int_0^\infty dt\,
  t^{\gamma-\frac{5}{2}}
  (t-\mu_j(x))_-^{2}\\
  &=B\left(\gamma-\frac{3}{2},3\right) \int_{-\infty}^{+\infty} dx
  \sum_{j=1}^\infty
  \mu_j^{\gamma+\frac{1}{2}}(x)\\
  &= B\left(\gamma-\frac{3}{2},3\right) \int_{-\infty}^{+\infty} \mbox{{
      tr}\ } W_-^{\gamma+\frac{1}{2}}(x)\,dx\,.
\end{align*}
{}From~\eqref{Lgd} we obtain
\begin{align*}
  L^{\mbox{\footnotesize\upshape cl}}_{\gamma,1}&=
\frac{\Gamma(\gamma+1)}
  {2\pi^{1/2}\Gamma(\gamma+\frac{3}{2})} = \frac{3}{16} \cdot
  \frac{\Gamma(\gamma+1)\Gamma(3)}
  {\Gamma(\gamma+\frac{3}{2})\Gamma(\frac{5}{2})} =\frac{3}{16}\cdot
  \frac{B\left(\gamma-\frac{3}{2},3\right)}
  {B\left(\gamma-\frac{3}{2},\frac{5}{2}\right)}\,,
\end{align*}
and this implies

\begin{theorem}\label{t2} 
Let the non-positive
operator family $W(x)$ satisfy~\eqref{WW}. Then
\begin{equation}
  \label{eq:LTHgamma}
  \tr\left(-\frac{d^2}{dx^2}\otimes {\boldsymbol 1}_{\boldsymbol G}+
W(x)
\right)_-^\gamma \leq L^{\mbox{\footnotesize\upshape cl}}_{\gamma,1}
\int_{-\infty}^{+\infty}\tr W^{\gamma+\frac{1}{2}}_-(x)\,dx\,.
\end{equation}
\end{theorem}

It remains to note, that the constant $L^{\mbox{\footnotesize\upshape
cl}}_{\gamma,1}$
in~\eqref{eq:LTHgamma}
is approached for
potentials $\alpha W$ as $\alpha\to+\infty$.

\section{Lieb-Thirring estimates with sharp constants for Schr\"odinger
operators in higher dimensions}

\subsection{Lieb-Thirring estimates for Schr\"odinger operators}

Let ${\boldsymbol G}$ be a separable Hilbert space. We
consider the operator
\begin{equation}
  \label{Sch0}
  -\Delta\otimes{\boldsymbol 1}_{\boldsymbol G}+V(x)
\end{equation}
in $L^2({\Bbb R}^d,{\boldsymbol G})$. If the family 
\begin{equation*}
V(\cdot) : {\Bbb R}^d\to {\mathscr B}
\end{equation*}
of bounded self-adjoint operators on ${\boldsymbol G}$ satisfies
\begin{equation}\label{d>d}
\|V(\cdot)\|_{\mathscr B} \in L^p({\Bbb R}^d),
\qquad \max\{1,d/2\}<p\leq\infty\,,
\end{equation}
then the quadratic form
\begin{equation*}
v[u,u]=\int_{{\Bbb R}^d} \left<V(x)u(x),u(x)\right>_{\boldsymbol G}\,dx
\end{equation*}
is zero-bounded with respect to
\begin{equation*}
h[u,u]=\int_{{\Bbb R}^d} \sum_{j=1}^d
\left\|\frac{\partial u}{\partial x_j}\right\|^2_{\boldsymbol G}\,dx\,.
\end{equation*}
This immediately follows  from 
the corresponding scalar result
and the arguments given when proving the 
inequalities~\eqref{eq:16},~\eqref{eq:17}.
Hence the quadratic form $h[\cdot,\cdot]+v[\cdot,\cdot]$ is semi-bounded
from below, closed on the Sobolev space $H^1({\Bbb R}^d,{\boldsymbol
G})$
and thus generates the operator~\eqref{Sch0}. As in subsection 3.2 one
can
show, that if in addition to~\eqref{d>d} we have $V(x)\in{\mathscr K}$
for a.e. $x\in{\Bbb R}^d$, then the negative spectrum of the 
operator~\eqref{Sch0} is discrete.

The main result of this paper is

\begin{theorem}\label{t3} 
  Assume that $V(x)\leq 0$ for a.e. $x\in{\Bbb R}^d$
and that $\tr V^{\frac{d}{2}+\gamma}(\cdot)$ is integrable 
for some $\gamma\geq 3/2$.  Then 
\begin{equation}
  \label{eq:LTHcl}
  \tr\left(-\Delta\otimes{\boldsymbol 1}_{\boldsymbol
G}+V(x)\right)_-^\gamma 
\leq L^{\mbox{\footnotesize\upshape cl}}_{\gamma,d} 
\int_{{\Bbb R}^d} \tr V^{\frac{d}{2}+\gamma}_-(x)\,dx\,.
\end{equation}
\end{theorem}

\noindent
{\bf Proof.} We use the induction arguments  
with respect to $d$.  For $d=1$, $\gamma\geq 3/2$
the bound~\eqref{eq:LTHcl} is identical to~\eqref{eq:LTHgamma}.
Assume that we have \eqref{eq:LTHcl} for  $d-1$ and all $\gamma\geq
3/2$. 
Consider the operator~\eqref{Sch0} 
in the (external) dimension $d$. 
We rewrite the quadratic form $h[u,u]+v[u,u]$ 
for $u\in H^1({\Bbb R}^d,{\boldsymbol G})$ as follows
\begin{align*}
h[u,u]+v[u,u]&=\int_{-\infty}^{+\infty} h(x_d)[u,u]\,dx_d
+\int_{-\infty}^{+\infty} w(x_d)[u,u]\,dx_d\,,\\
h(x_d)[u,u]&=\int_{{\Bbb R}^{d-1}}
\left\|\frac{\partial u}{\partial x_d}\right\|^2_{\boldsymbol
G}dx_1\cdots x_{d-1}\,,\\
w(x_d)[u,u]&=\int_{{\Bbb R}^{d-1}}
\left[
\sum_{j=1}^{d-1}\left\|\frac{\partial u}{\partial
x_j}\right\|^2_{\boldsymbol G}
+
\left<V(x)u,u\right>_{\boldsymbol G}
\right]
dx_1\cdots x_{d-1}\,.
\end{align*}
The form $w(x_d)$ is closed on $H^1({\Bbb R}^{d-1},{\boldsymbol G})$ for
a.e.
$x_d\in{\Bbb R}$ and 
it induces the self-adjoint operator
\begin{equation*}
W(x_d)=-\sum_{k=1}^{d-1}\frac{\partial^2}{\partial x_k^2}\otimes
{\boldsymbol 1}_{\boldsymbol G}
+V(x_1,\dots,x_{d-1};x_d) 
\end{equation*}
on $L^2({\Bbb R}^{d-1},{\boldsymbol G})$. The negative spectrum of this 
$(d-1)$--dimensional Schr\"odinger system
is discrete, hence $W_-(x_d)$ is compact
on $L^2({\Bbb R}^{d-1},{\boldsymbol G})$
and according to our induction hypothesis
$\tr W_-^{\gamma+\frac{1}{2}}(x_d)$ satisfies the inequality
\begin{equation}\label{eq:trwww}
\tr W^{\gamma+\frac{1}{2}}_-(x_d)
\leq L^{\mbox{\footnotesize\upshape
cl}}_{\gamma+\frac{1}{2},d-1}\int_{{\Bbb R}^{d-1}}
\tr V_-^{\gamma+\frac{d}{2}}(x_1,\dots,x_{d-1};x_d)
\,dx_1\cdots dx_{d-1}
\end{equation}
for a.e. $x_d\in{\Bbb R}$. For $V\in L^{\gamma+\frac{d}{2}}({\Bbb
R}^{d-1})$,
the  function $\tr W_-^{\gamma+\frac{1}{2}}(\cdot)$ is
integrable. 

Let $w_-(x_d)[\cdot,\cdot]$ be the quadratic form corresponding to
the operator $W_-(x_d)$ on ${\boldsymbol H}=L^2({\Bbb
R}^{d-1},{\boldsymbol G})$.
Then we have $w(x_d)[u,u]\geq -w_-(x_d)[u,u]$ and
\begin{equation}\label{eq:rhs}
h[u,u]+v[u,u]\geq \int_{-\infty}^{+\infty}
\left[
\left\|\frac{\partial u}{\partial x_d}\right\|^2_{\boldsymbol H}
-\left<W_-(x_d)u,u\right>_{\boldsymbol H}
\right]\,dx_d
\end{equation}
for all $u\in H^1({\Bbb R}^{d},{\boldsymbol G})$. 
According to section 2.2 the form on the r.h.s.\ of~\eqref{eq:rhs}
can be closed to
$H^1({\Bbb R},{\boldsymbol H})$ and induces the self-adjoint operator
\begin{equation*}
-\frac{d^2}{dx_d^2}\otimes{\boldsymbol 1}_{\boldsymbol H} -
W_-(x_d)
\end{equation*}
on $L^2({\Bbb R},{\boldsymbol H})$.
Then~\eqref{eq:rhs} implies
\begin{equation}\label{eq:llaass}
  \tr(-\Delta\otimes
{\boldsymbol 1}_{\boldsymbol G}+V)_-^\gamma \leq \tr\left(
    -\frac{d^2}{dx_d^2}\otimes{\boldsymbol 1}_{\boldsymbol H} -
W_-(x_d)\right)_-^\gamma \,.
\end{equation}
We can now apply~\eqref{eq:LTHgamma} to the r.h.s. of~\eqref{eq:llaass}
and
in view of~\eqref{eq:trwww} we find
\begin{align*}
\tr\left(
    -\frac{d^2}{dx_d^2}\otimes{\boldsymbol 1}_{\boldsymbol H} -
    W_-(x_d)\right)_-^\gamma
  &\leq L^{\mbox{\footnotesize\upshape
cl}}_{\gamma,1}\int_{-\infty}^{+\infty}\
    \tr W_-^{\gamma+\frac{1}{2}}(x_d)\,dx_d\\
  &\leq L^{\mbox{\footnotesize\upshape
cl}}_{\gamma,1}L^{\mbox{\footnotesize\upshape
cl}}_{\gamma+\frac{1}{2},d-1} \int_{{\Bbb
      R}^d} \tr V_-^{\gamma+\frac{d}{2}}(x)\,dx\,.
\end{align*}
The calculation
\begin{align*}
  L^{\mbox{\footnotesize\upshape
cl}}_{\gamma,1}L^{\mbox{\footnotesize\upshape
cl}}_{\gamma+\frac{1}{2},d-1}
  &= \frac{\Gamma(\gamma+1)}
{2\pi^{\frac{1}{2}}\Gamma(\gamma+\frac{1}{2}+1)}
  \cdot \frac{\Gamma(\gamma+\frac{1}{2}+1)}
  {2^{d-1}\pi^{\frac{d-1}{2}}
  \Gamma(\gamma+ \frac{1}{2}+\frac{d-1}{2}+1)}\\
  &=
\frac{\Gamma(\gamma+1)}{2^d\pi^{\frac{d}{2}}\Gamma(\gamma+\frac{d}{2}+1)}
  =L^{\mbox{\footnotesize\upshape cl}}_{\gamma,d}
\end{align*}
completes the proof.  \vspace{2mm}

For the special case ${\boldsymbol G}={\Bbb C}$ we obtain the
Lieb-Thirring
bounds
for scalar Schr\"odinger operators with the (sharp) classical constant
$L_{\gamma,d}=L^{\mbox{\footnotesize\upshape cl}}_{\gamma,d}$ 
for $\gamma\geq 3/2$ in all 
dimensions $d$.

\subsection{Lieb-Thirring estimates for magnetic operators}

Following a remark by B. Helffer \cite{Hel} we demonstrate, how 
Theorem~\ref{t3} can be extended to Schr\"odinger operators with magnetic
fields. Let 
\begin{equation*}
{\boldsymbol a}(x)=(a_1(x),\dots,a_d(x))^t,
\qquad  d\geq 2\,,
\end{equation*}
be a magnetic vector potential with real-valued entries 
$a_k\in L^2_{\footnotesize loc}({\Bbb R}^d)$. 
Put
\begin{equation*}
H({\boldsymbol a})=(i\nabla+{\boldsymbol a}(x))^2\otimes{\boldsymbol
  1}_{\boldsymbol G}\,.
\end{equation*}
Its form domain $d[h({\boldsymbol a})]$
consists of the closure of
all smooth compactly supported functions with respect to 
$h({\boldsymbol a})[\cdot,\cdot]
+\|\cdot\|^2_{L^2({\Bbb R}^d,{\boldsymbol G})}$ (cf. \cite{Si}), where
\begin{equation*}
h({\boldsymbol a})[u,u]=
\sum_{k=1}^d
\int_{{\Bbb R}^d}
\left\|\left(i\frac{\partial}{\partial
x_k}+a_k\right)u\right\|^2_{\boldsymbol G}\,dx\,.
\end{equation*}
Let the operator family $V$ and the corresponding form $v$ be defined
as in the previous subsection. If~\eqref{d>d} is satisfied, then one can
apply Kato's inequality \cite{Kato,Si}, 
and find that the form
\begin{equation}\label{qqvv}
q({\boldsymbol a})[u,u]=h({\boldsymbol a})[u,u]+v[u,u]
\end{equation}
is closed on $d[q({\boldsymbol a})]=d[h({\boldsymbol a})]$ and induces
the
self-adjoint operator
\begin{equation}\label{mSch0}
Q({\boldsymbol a})=
H({\boldsymbol a}) + V(x)
\end{equation}
on $L^2({\Bbb R}^d,{\boldsymbol G})$. Finally, by applying Kato's
inequality to the higher-dimensional analog of~\eqref{eq:16}
we see, that $V(x)\in{\mathscr K}$ for a.e. $x\in{\Bbb R}^d$
in conjunction with~\eqref{d>d}
implies that $Q({\boldsymbol a})$ has discrete negative
spectrum.

\begin{theorem}\label{mt3} 
  Assume that 
  {\upshape ${\boldsymbol a}\in L^2_{\footnotesize loc}({\Bbb R}^d)$ }
  is a real vector field,
  and that the non-positive operator family $V(x)$ satisfies
  $\tr V^{\frac{d}{2}+\gamma}\in L^1({\Bbb R}^d)$ for some
  $\gamma\geq 3/2$.  Then 
\begin{equation}
  \label{eq:mLTHcl}
  \tr\left(H({\boldsymbol a})+V(x)\right)^\gamma_- 
\leq L^{\mbox{\footnotesize\upshape cl}}_{\gamma,d} 
\int_{{\Bbb R}^d} \tr V^{\frac{d}{2}+\gamma}_-\,dx\,.
\end{equation}
\end{theorem}

\noindent
{\bf Proof.}  
In the dimension $d=1$, any magnetic field can be removed
by gauge transformation. Thus~\eqref{eq:LTHgamma} can serve to
initiate the induction procedure.

Assume now that~\eqref{eq:mLTHcl} is known for 
all $\gamma\geq 3/2$ for the dimension
$d-1$ and consider the operator $H({\boldsymbol a})$
in the dimension $d$.
Put
\begin{equation*}
W(x_d)=
\left[\sum_{n=1}^{d-1}
\left(i\frac{\partial}{\partial
x_n}+a_n(x)\right)^2\right]
+V(x)\,.
\end{equation*}
We find that 
\begin{align*}
q({\boldsymbol a})[u,u]&=
\int_{{\Bbb R}^d} 
\left\|\left(i\frac{\partial}{\partial
x_d}+a_d\right)u\right\|^2_{\boldsymbol G}dx
+
\int_{{\Bbb R}}\left<W(x_d)u,u\right>_{{\boldsymbol H}}
dx_d\\
&\geq
\int_{{\Bbb R}^d} 
\left\|\left(i\frac{\partial}{\partial
x_d}+a_d\right)u\right\|^2_{\boldsymbol G}dx
-
\int_{{\Bbb R}}\left<W_-(x_d)u,u\right>_{{\boldsymbol H}}
dx_d\,,
\end{align*}
where for fixed $x_d\in{\Bbb R}$ the operator $W_-(x_d)$ 
is the negative part of
$W(x_d)$ on ${\boldsymbol H}=L^2({\Bbb R}^{d-1},{\boldsymbol G})$.
We now choose a gauge in which $a_d$ vanishes. Namely,
put
\begin{equation*}
\phi(x_1,\dots,x_d)=\int_0^{x_d}a_d(x_1,\dots,x_{d-1},\tau)d\tau
\end{equation*}
and $\tilde{u}(x)=e^{i\phi(x)}u(x)$ for all $u\in d[q({\boldsymbol
a})]$.
Then
\begin{equation}\label{eq:dd}
q({\boldsymbol a})[u,u]\geq
\int_{{\Bbb R}^d} 
\left\|\frac{\partial \tilde{u}}{\partial x_d}\right\|^2_{\boldsymbol
G}dx
-
\int_{{\Bbb
R}}\left<\tilde{W}(x_d)\tilde{u},\tilde{u}\right>_{{\boldsymbol H}}
dx_d\,,
\qquad u\in d[q({\boldsymbol a})]\,,
\end{equation}
where
\begin{equation*}
\widetilde{W}(x_d)=e^{i\phi(x^\prime,x_d)}W_-(x_d)e^{-i\phi(x^\prime,x_d)}\,,
\qquad x^\prime=(x_1,\dots,x_{d-1})\,,
\end{equation*}
acts on ${\boldsymbol H}$ for any fixed $x_d\in{\Bbb R}$.
Closing the form on the r.h.s.\ of~\eqref{eq:dd} we see that
\begin{equation}\label{aaaaa}
\tr\left(H({\boldsymbol a})+V(x)\right)^\gamma_- 
  \leq 
\tr\left( 
-\frac{d^2}{dx_d^2}\otimes{\boldsymbol 1}_{\boldsymbol H} 
- \widetilde{W}(x_d)\right)_-^\gamma\,,
\end{equation}
where the
operator on the r.h.s.\ acts in $L^2({\Bbb R},{\boldsymbol H})$.
{}From our induction hypothesis we have
\begin{align*}
\tr \widetilde{W}^{\gamma+\frac{1}{2}}(x_d)
=\tr W^{\gamma+\frac{1}{2}}_-(x_d)
\leq L^{\mbox{\footnotesize\upshape cl}}_{\gamma+\frac{1}{2},d-1}
\int_{{\Bbb R}^{d-1}}
\tr V_-^{\gamma+\frac{d}{2}}(x^\prime;x_d)\,dx^\prime\,.
\end{align*}
Hence~\eqref{eq:LTHgamma}
can be applied to estimate the r.h.s.\ of \eqref{aaaaa}
and we complete the proof of~\eqref{eq:mLTHcl} 
in the same manner as in the proof of Theorem~\ref{t3}.

\subsection{Lieb-Thirring estimates for the Pauli operator}

As an application of Theorem~\ref{mt3} we deduce a Lieb-Thirring type
bound
for the Pauli operator. Preserving the notations of the previous
subsection we
put $d=3$ and ${\boldsymbol G}={\Bbb C}^2$.  
Let ${\boldsymbol a}(x)=(a_1(x),a_2(x),a_3(x))^t$ be a twice
continuously differentiable vector function with real-valued entries.
The Pauli operator is given by the differential expression
\begin{equation}\label{Pauli}
Z=Q({\boldsymbol a})\otimes{\boldsymbol 1} 
+
\begin{pmatrix}
  a_{1,2} & -ia_{3,1}+a_{2,3}\\
  ia_{3,1}+a_{2,3} & - a_{1,3}
\end{pmatrix}
+
V\otimes{\boldsymbol 1},
\end{equation}
where ${\boldsymbol 1}$ is the identity on ${\Bbb C}^2$, $V=V(x)$ is the
multiplication by a real-valued scalar potential and
\begin{equation*}
a_{j,k}=\frac{\partial a_j}{\partial x_k}-
\frac{\partial a_k}{\partial x_j}, \qquad k,j=1,2,3\,.
\end{equation*}
Let $B(x)$ be the length of the vector 
${\mathscr B}(x)=\mbox{{ curl}\ } {\boldsymbol a}(x)$.  Then
the two eigenvalues of the perturbation of the term 
$Q({\boldsymbol a})\otimes{\boldsymbol 1}$ in
\eqref{Pauli} at some point $x\in{\Bbb R}^3$ are given by
\begin{equation*}V(x)\pm B(x).\end{equation*}
If $V,B\in
L^{\gamma+\frac{3}{2}}({\Bbb R}^3)$ for some $\gamma\geq 3/2$, then
Theorem~\ref{mt3} 
implies
\begin{equation}\label{LTPauli}
\tr Z_-^\gamma
\leq L^{\mbox{\footnotesize\upshape cl}}_{\gamma,3}
\left(
\int \left\{(V+B)_-^{\gamma+\frac{3}{2}}
+(V-B)_-^{\gamma+\frac{3}{2}} \right\}
dx
\right).
\end{equation}

\subsection{Acknowledgements}
The authors wish to express their 
gratitude to V.S. Buslaev and L.D. Faddeev 
for the useful discussions concerning trace formulae
and B. Helffer for his valuable comments on
magnetic Schr\"odinger operators. 
The first author has been supported by the Swedish Natural 
Sciences Research Council, 
Grant M-AA/MA 09364-320.
The second author has been supported by the Swedish Natural Science
Council dnr 11017-303.
Partial financial support from
the European Union through the TMR network FMRX-CT 96-0001 is
gratefully acknowledged.

{\small
\begin{multicols}{2}
  \raggedright
  Royal Institute of Technology$^1$\\
  Department of Mathematics\\
  S-10044 Stockholm, Sweden\\
  Universit\"at Regensburg$^2$\\
  Naturwissenschaftliche Fakult\"at I\\
  D-93040 Regensburg, Germany
\end{multicols}
} }


\begin{thebibliography}{99}

%
\bibitem{AL} Aizenmann M. and Lieb E.H.: On semi-classical bounds for
eigenvalues of Schr\"odinger operators. Phys. Lett. {\bf 66A}, 427-429
(1978)
%
%
\bibitem{Bz}Berezin F.A.:  
Covariant and contravariant symbols of operators. 
[English translation], Math. USSA Izv. 
 {\bf 6}, 1117-1151 (1972)
%
%
\bibitem{Bla} Blanchard Ph.\ and Stubbe J.:
Bound states for Schr\"odinger Hamiltonians: Phase Space Methods
and Applications. Rev. Math. Phys., {\bf 35}, 504-547 (1996)
%
%
\bibitem{B} De la Bret\`eche R.:  
Preuve de la conjecture de Lieb-Thirring dans
le cas des potentiels quadratiques strictement convexes.
 to be published in Ann. Inst. Henri Poincar\'e
Physique th\'eorique. 
%
%
\bibitem{BF} Buslaev V.S. and Faddeev L.D: 
Formulas for traces for a singular Sturm-Liouville differential 
operator. [English translation],
 Dokl. AN SSSR, {\bf 132}, 451-454(1960)
%
%
\bibitem{Con} Conlon J.G.: A new proof of the Cwikel-Lieb-Rosenbljum
bound. Rocky Mountain J. Math., {\bf 15}, 117-122 (1985)
%
%
\bibitem{C} Cwikel M.: Weak type estimates for singular values and the
number of bound states of Schr\"odinger operators. Trans. AMS, {\bf
224},
93-100
  (1977)
%
%
\bibitem{FadZ} Faddeev L.D. and Zakharov, V.E.: Korteweg-de Vries
equation:
A completely integrable hamiltonian system. Func. Anal. Appl., {\bf 5}, 
18-27  (1971)
%
%
\bibitem{GGM} Glaser V., Grosse H. and Martin A.:  
Bounds on the number of eigenvalues of the Schr\"odinger operator. 
Commun. Math. Phys., {\bf 59}, 197-212 (1978)
%
%
\bibitem{Hel} Helffer B.: private communication
%
%
\bibitem{RoHe} Helffer B. and Robert, D.: Riesz means of bounded states
and
  semi-classical limit connected with a Lieb-Thirring conjecture I,II. 
I - Jour. Asymp. Anal., {\bf 3}, 91-103 (1990), 
II - Ann. de l'Inst. H. Poincare, {\bf 53} (2), 139-147 (1990)
%
%
\bibitem{HLW} Hundertmark D., Laptev A. and Weidl T.: 
New  bounds on the Lieb-Thirring constants. Preprint KTH, Sweden.
%
%
\bibitem{HLT} Hundertmark D., Lieb E.H. and Thomas L.E.: A sharp bound
for an
eigenvalue moment of the one-dimensional Schr\"odinger operator.  
Adv.\ Theor.\ Math. Phys.\ \textbf{2}, 719-731 (1998)
%
%
\bibitem{Kato} Kato T.: Schr\"odinger operators with singular
potentials.
Israel J. Math., {\bf 13} 135-148 (1973)
%
\bibitem{Lap1} Laptev A.: Dirichlet and Neumann Eigenvalue Problems 
on Domains in Euclidean Spaces. J. Func. Anal., {\bf 151}, 
531-545  (1997)
%
%
\bibitem{Lap2} Laptev A.: 
On inequalities for the bound states of Schr\"odinger operators.
Operator Theory: Advances and Applicatons, 
Birkh\"auser Verlag Basel/Switzerland, {\bf 78}, 221-225 
(1995)
%
%
\bibitem{LS} Laptev A. and Safarov Yu.:
A generalization of the Berezin-Lieb inequality. 
Amer. Math. Soc. Transl (2)
{\bf 175}, 69-79 (1996)
%
%
\bibitem{LY} Li P. and Yau S.-T.: On the Schr\"odinger equation and 
the eigenvalue problem.
Comm. Math. Phys.,  {\bf 88},
309-318 (1983)
%
\bibitem{L1} Lieb E.H.: The number of bound states of one body
Schr\"odinger operators and the Weyl problem.  Bull. Amer. Math. Soc. 
{\bf 82}, 751-753 (1976)
%
%
\bibitem{L2} Lieb E.H.: The classical limit of quantum
spin systems. Comm. Math. Phys.
{\bf 31}, 327-340  (1973)
%
%
\bibitem{L3} Lieb E.H.: On characteristic exponents in turbulence.  
Comm. Math. Phys., {\bf 82}, 473-480 (1984)
%
%
\bibitem{LT} Lieb E.H. and Thirring, W.: Inequalities for the moments of
the
  eigenvalues of the Schr\"odinger Hamiltonian and their relation to
  Sobolev inequalities.  Studies in Math. Phys., Essays in Honor of
Valentine
  Bargmann., Princeton, 269-303 (1976)
%
%
\bibitem{P} P\'olya G.: On the eigenvalues of vibrating membranes. 
 Proc. London Math. Soc. {\bf 11}, 
419-433  (1961)
%
%
\bibitem{R} Rozenblum G.V.: Distribution of the discrete spectrum of
singular differential operators. Dokl. AN SSSR, {\bf 202},  1012-1015
(1972), Izv.  VUZov, Matematika, {\bf 1}, 75-86 (1976)
%
\bibitem{Si} Simon B.: Maximal and minimal Schr\"odinger Forms.
J. Operator Theory {\bf 1} 37-47 (1979)
%
\bibitem{W1} Weidl T.: On the Lieb-Thirring constants $L_{\gamma,1}$
for $\gamma\geq 1/2.$ Comm. Math. Phys., {\bf 178}, 135-146 (1996)
%
\end{thebibliography}
\end{document}